\documentstyle[prl,aps,epsf,twocolumn]{revtex}
\begin{document}
\draft
\title{High temperature anomaly of the conductance
of a tunnel junction}
\author{Georg G\"oppert, Xiaohui Wang, and Hermann Grabert}
\address{Fakult{\"a}t f{\"u}r Physik, 
Albert-Ludwigs-Universit{\"a}t Freiburg,\\ 
Hermann-Herder-Stra{\ss}e  3, D-79104  Freiburg}
\maketitle
\begin{abstract}
The linear conductance of a tunnel junction in series with an 
ohmic resistor is determined in the high temperature limit. 
The tunneling current is treated nonperturbatively by means 
of path integral techniques. Due to quantum effects the 
conductance is smaller than the classical series conductance. 
The reduction factor is found to be nonanalytic in the 
environmental resistance for vanishing resistance. 
This behavior is a high temperature manifestation of 
the Coulomb blockade effect.
\end{abstract}

\narrowtext
In the last decade or so tremendous progress was made in our 
understanding and control of Coulomb blockade phenomena 
\cite{GD,Kastner}. 
These effects arise in nanostructures with small capacitances 
$C$ when the single electron charging energy $E_C=e^2/2C$ 
becomes significant. So far work has concentrated on the 
low-temperature, typically millikelvin region to satisfy 
the seemingly obvious condition $k_BT \ll E_C$ for the 
manifestation of charging effects. However, recently 
Pekola and coworkers \cite{Pekola} have demonstrated that 
there are remarkable signatures of the Coulomb blockade 
effect also for temperatures in the few Kelvin region well 
above the charging energy. In fact, charging phenomena in 
the high-temperature region can be used for precision 
thermometry \cite{Pekola}.

Further, in most of the theoretical studies of charging effects 
in metallic nanostructures it is assumed that the tunneling 
resistance $R_T$ of the tunnel junctions in the device exceeds 
the von-Klitzing resistance $R_K=h/e^2 \approx 25.8 k \Omega$. 
Under the condition $R_T \gg R_K$ quantum fluctuations of the 
charge \cite{AN} are strongly suppressed. Hence, in a sense 
this latter condition ensures that charging effects are not 
washed out by quantum fluctuations in much the same way as 
the above-mentioned temperature constraint ensures 
stability against thermal fluctuations.

It is now interesting to examine whether signatures of the 
Coulomb blockade effect  persist when thermal fluctuations 
are large and the tunneling resistance may become small. 
This question will be addressed for the fundamental problem 
of an ultrasmall tunnel junction biased by a voltage source. 
The single junction system was studied previously 
\cite{Devoret} in the limit $R_T \gg R_K$ where 
tunneling can be treated perturbatively. While the theoretical 
approach employed  in these earlier studies does not readily 
extend to the case of strong tunneling, we know from these 
investigations that the impedance of the leads connecting 
the tunnel junction with the voltage source crucially affects 
the measurable conductance. In fact, charging effects in single 
junctions only arise from the coupling of the tunneling 
electrons to the modes of the electromagnetic environment 
of the junction \cite{IN}. In the model studied in this 
work the environmental impedance will be assumed to be ohmic.
\begin{figure}
\epsfysize=9cm
\epsffile{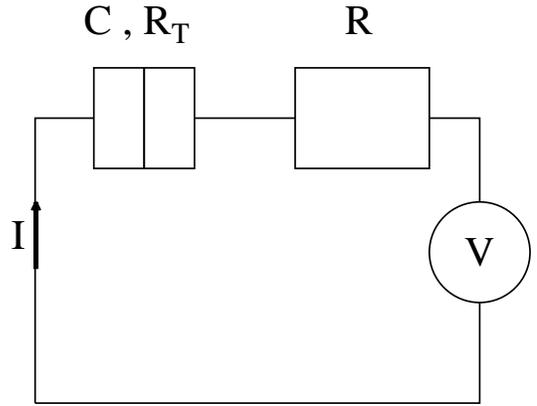}
\caption[]{\label{fig1} Circuit diagram of a tunnel junction with
capacitance $C$ and tunneling resistance $R_T$ biased by a
voltage source $V$ via leads of impedance $R$.} 
\end{figure}
Specifically, we consider the circuit depicted in Fig.~\ref{fig1}. 
We aim at the zero-bias differential conductance 
$G= \partial I/ \partial V|_{V=0}$ where $I$ is the current and 
$V$ the applied voltage. Our approach is based on the path integral 
formulation of the tunnel junction \cite{BMS} and employs the 
generating functional
\begin{equation}
Z[\xi] =  {\rm tr} \, T_\tau \exp \left\{ - \frac{1}{\hbar} 
\int^{\hbar \beta}_0 d \tau \, [H-I \xi(\tau)] \right\} 
\label{func}
\end{equation}
where $H$ is the Hamiltonian of the system for $V=0$ and $I$ 
the microscopic current operator for charge flow in the leads. 
$T_{\tau}$ is the time ordering operator for the imaginary-time 
variable $\tau$.  Introducing a phase variable $\varphi$ 
\cite{BMS} which is related to the voltage across the 
tunnel junction via a Josephson-type relation, the generating 
functional (\ref{func}) may be written as a path integral
\begin{equation}
Z[\xi]  =  \int D[\varphi] \exp  
\left\{- \frac{1}{\hbar} \, S [\varphi, \xi] \right\} \label{pathint}
\end{equation}
with the effective Euclidean action 
\begin{equation}
 S[\varphi,\xi]  =  S_C[\varphi] +S_T[\varphi]  
+ S_R[\varphi, \xi] \, . \label{action}
\end{equation}
where
\[
S_C[\varphi] = \int^{\hbar \beta}_0  d \tau \, 
\frac{\hbar^2C}{2e^2} \, \dot{\varphi}^2 
\]
describes Coulomb charging of the junction and 
\begin{equation}
S_T[\varphi] = 2 \int^{\hbar \beta}_0 \! d \tau  
\int^{\hbar \beta}_0 \! d \tau^\prime \, \alpha(\tau 
- \tau^\prime) \sin^2 \! \left[\frac{\varphi(\tau)-
\varphi(\tau^\prime)}{2} \right] \label{tunnel}
\end{equation}
quasi-particle tunneling across the junction \cite{BMS}. 
The kernel $\alpha(\tau)$ is determined by the tunneling 
resistance $R_T$ and may be written as 
\[
\alpha(\tau)  = \frac{1}{\hbar \beta} \sum^{+ 
\infty}_{n=- \infty}  \widetilde{\alpha}(\nu_n) 
\, e^{-i \nu_n \tau}
\]
where the $\nu_n = 2 \pi n/ \hbar \beta$ are Matsubara 
frequencies and
\[
\widetilde{\alpha}(\nu_n) = - \frac{\hbar}{4 \pi} \, 
\frac{R_K}{R_T} \left| \nu_n \right|  \, .
\]
The last term in the action (\ref{action}) reads
\begin{eqnarray}
S_R[\varphi, \xi] &=& \frac{1}{2} \int^{\hbar 
\beta}_0  d \tau  \int^{\hbar \beta}_0  d \tau^\prime 
k(\tau - \tau^\prime) \nonumber\\
&& \times \left[\varphi(\tau) + \frac{e}{\hbar}\xi(\tau) -
\varphi(\tau^\prime ) - \frac{e}{\hbar} \xi(\tau^\prime)\right]^2
\label{resistor}
\end{eqnarray}
For $\xi(\tau) \equiv 0$ this is the well-known influence 
functional of an Ohmic lead resistance $R$ \cite{CL} with the kernel 
\begin{equation}
k(\tau) = \frac{1}{\hbar \beta} \sum^{+ \infty}_{n=- \infty} 
\widetilde{k} (\nu_n) \, e^{-i \nu_n \tau} \label{kernel}
\end{equation}
where 
\[
\widetilde{k} (\nu_n) = - \frac{\hbar}{4 \pi} \, \frac{R_K}{R} 
\left| \nu_n \right| \, . 
\]
Note that the influence functional (\ref{resistor}) of the 
lead resistance is at most quadratic in the phase while the 
tunneling term (\ref{tunnel}) with a trigonometric phase
dependence renders the path integral 
(\ref{pathint}) non-Gaussian. This is  due to the discrete charge 
transfer across the tunnel junction.

The conductance $G$ can be calculated from the Kubo formula
\begin{equation}
G = \lim_{\omega \rightarrow 0} \frac{1}{\hbar \omega } {\rm Im} 
\left[ \lim_{i \nu_n \rightarrow \omega  + i \delta} 
\int^{\hbar \beta}_0 \! d \tau\, e^{i \nu_n \tau} 
\langle I( \tau) I(0)\rangle \right] \label{conduct}
\end{equation}
where the current-current correlator $\langle I(\tau)I(0)\rangle$ reads 
in terms of the generating functional (\ref{func})
\begin{equation}
\langle I(\tau)I(0)\rangle = \frac{\hbar^2}{Z[0]} \, \left. 
\frac{\delta^2Z[\xi]}{\delta \xi(\tau) \delta \xi (0)} 
\right|_{\xi(\tau) \equiv 0} \, . \label{correlator}
\end{equation}
Now, using the path integral representation (\ref{pathint}) 
and Eq.~(\ref{resistor}), the correlator may be written as
\begin{eqnarray}
\langle I(\tau)I(0)\rangle &=& \frac{1}{Z}  \int 
D[\varphi] 
\exp \left\{ - \frac{1}{\hbar} S[\varphi] \right\} 
\label{phicorr} \\
&&\times
\left( 2 \frac{e^2}{\hbar} k(\tau) + I [\varphi, \tau] 
I [\varphi, 0] \right) \nonumber
\end{eqnarray}
where $S[\varphi]=S[\varphi, \xi \equiv 0]$ and 
\begin{equation}
Z = \int D[\varphi]  \exp\left\{- \frac{1}{\hbar} 
S[\varphi]\right\}
\label{part}
\end{equation}
is the partition function. Further, the current functional 
$I[\varphi, \tau]$ is given by 
\begin{equation}
I [\varphi, \tau] = \frac{2e}{\hbar} 
\int^{\hbar \beta}_0 d \tau^\prime \,  k(\tau - \tau^\prime) 
\varphi(\tau^\prime) \, .
\label{currfunc}
\end{equation}
To calculate the conductance from Eq.~(\ref{conduct}), 
we first need to determine the Fourier components
\begin{equation}
C(\nu_n) = \int^{\hbar \beta}_0  d \tau \,  e^{i \nu_n \tau} 
\langle I(\tau) I(0)\rangle \, . \label{ctot}
\end{equation}
Using Eqs.~(\ref{kernel}), (\ref{phicorr}), and 
(\ref{currfunc}), one finds that 
$C(\nu_n)=C_1(\nu_n)+C_2(\nu_n)$ where
\begin{equation}
C_1(\nu_n) = 2 \frac{e^2}{\hbar} \, \widetilde{k} (\nu_n) 
\label{cone}
\end{equation}
and
\begin{equation}
C_2(\nu_n) = \frac{1}{Z} \, \int D[\varphi] 
\exp \left\{- \frac{1}{\hbar} S [\varphi]\right\} F[\varphi, \nu_n] \, . 
\label{ctwo}
\end{equation}
Here the functional
\begin{equation}
F[\varphi, \nu_n] = \frac{4e^2 \beta}{\hbar} \, 
\widetilde{k}(\nu_n) \widetilde{\varphi}(\nu_n)  
\sum^{+ \infty}_{m=- \infty}  \widetilde{k}(\nu_m) 
 \widetilde{\varphi} (\nu_m) \label{ctwofunc}
\end{equation}
is given in terms of the Fourier coefficients 
$\widetilde{\varphi}(\nu_n)$ of the phase variable. 
Since the action $S[\varphi]$ is invariant 
under a global phase shift, we may put $\widetilde\varphi(0)=0$.

According to the decomposition of $C(\nu_n)$ into (\ref{cone}) 
and (\ref{ctwo}), the conductance (\ref{conduct}) splits into 
$G=G_1+G_2$ where
\begin{equation}
G_1 = \lim_{\omega  \rightarrow 0} \, \frac{1}{\hbar \omega } \, {\rm Im} 
\, \frac{2e^2}{\hbar} \widetilde{k} (-i\omega  + \delta) 
= \frac{1}{R} \, . \label{gone}
\end{equation}
Since the kernel $k(\tau)$ has the analytic properties of 
an equilibrium correlation function, the analytic 
continuation of $\widetilde{k}(\nu_n)$ is unique \cite{BM}.

To proceed we first note that in terms of the Fourier coefficients 
$\tilde{\varphi}(\nu_n)$ the action functional may be written
\[
S[\varphi]=S_0[\varphi]+\sum_{k=2}^{\infty} S_{2k}[\varphi]
\] 
where 
\begin{equation}
S_0[\varphi]= \hbar \sum_{n=1}^{\infty} \lambda(\nu_n) 
|\tilde{\varphi}(\nu_n)|^2 
\label{snull}
\end{equation}
with the eigenvalues 
\begin{equation}
\lambda(\nu_n)= \hbar \beta \left( \frac{\hbar \nu_n^2}{2 E_c}+ 
\frac{g}{2 \pi}|\nu_n| \right) \, . \label{lam}
\end{equation}
Here $g=R_K\left(R_T^{-1}+R_{}^{-1}\right)$
is the dimensionless parallel conductance of
the tunneling and lead resistances. Further, 
\begin{eqnarray*}
 S_{2k}[\varphi] &=&\frac{(-1)^{k+1}}{(2k)!} \sum_{l=1}^{2k-1} 
{2k \choose l} (-1)^l \hbar\beta \\ 
& & \times  \sum_{{n_1, \cdots, n_{2k-1}\atop{\{n_p\ne 0\}}}} 
 \tilde{\alpha}(-\sum\nolimits_{p=1}^{l}
\nu_{n_p}) \tilde{\varphi}(\nu_{n_1}) \cdots \\
 & & \times \
\tilde{\varphi}(\nu_{n_{2k-1}}) \tilde{\varphi}
(-\sum\nolimits_{p=1}^{2k-1} \nu_{n_p}). 
 \end{eqnarray*}
Now, in the high-temperature limit we may expand 
about the Gaussian action $S_0[\varphi]$ using
\begin{eqnarray}
\exp(- S[\varphi]/\hbar)
&=& \exp(-S_0[\varphi]/\hbar)
[ 1 - S_4[\varphi]/\hbar \label{exp}\\
&& -S_6[\varphi]/\hbar  - S_8[\varphi]/\hbar 
+S_4[\varphi]^2/2\hbar^2 + \cdots ] \, .
\nonumber
\end{eqnarray} 
The ratio of path integrals (\ref{ctwo}) can then
be evaluated by means of
\[
\frac{\int D[\varphi]
\exp\left\{-\frac{1}{\hbar} S_0[\varphi]\right\}
 \widetilde{\varphi}(\nu_k)\widetilde{\varphi}(\nu_l) }
{\int D[\varphi]
\exp\left\{-\frac{1}{\hbar} S_0[\varphi]\right\}}
= \frac{\delta_{k,-l}}{\lambda(\nu_k)} 
\]
and Wick's theorem for higher order products
of the Fourier coefficients $\widetilde\varphi(\nu_n)$.
Since
\[
\frac{1}{\lambda(\nu_k)}=\frac{\beta E_C}{2\pi^2k^2+g\beta E_C|k|}
\]
is of order $\beta E_C$, the expansion (\ref{exp})
gives a high-temperature series in powers of $\beta E_C$.

Proceeding along these lines, one obtains from
Eqs.~(\ref{ctwo}) and (\ref{ctwofunc})
\begin{eqnarray}
C_2(\nu_n) &=& \frac{4e^2\beta}{\hbar}\,
\frac{\widetilde k(\nu_n)^2}{\lambda(\nu_n)}
\Big( 1 \nonumber \\
&& + \frac{2\beta}{\lambda(\nu_n)}
\sum_{{m=-\infty}\atop {m\ne 0}}^{\infty}
\frac{\widetilde\alpha(\nu_{n+m})-\widetilde\alpha(\nu_n)
-\widetilde\alpha(\nu_m)}{\lambda(\nu_m)}
\nonumber\\
&&+ \ {\cal O}\left(\beta E_C\right)^3
\Big)\, . \label{ctwoex}
\end{eqnarray}
In view of Eqs.~(\ref{conduct}) 
and (\ref{ctot}), the high-temperature series for 
$C_2(\nu_n)$ yields a corresponding expansion of the conductance 
\begin{equation}
G_2 = \lim_{\omega \rightarrow 0} \frac{1}{\hbar \omega} 
{\rm Im} \, C_2(-i \omega + \delta) \, . \label{gtwo}
\end{equation}
However, the analytic continuation and the limit 
$\omega \rightarrow 0$ involve a subtlety. 
As is readily seen from Eq.~(\ref{lam}), a factor 
$1/ \lambda(\nu_n)$ in Eq.~(\ref{ctwoex}) gives in this limit 
a factor $2 \pi i/g \beta \hbar \omega$. While the 
$1/ \omega$-divergences  are canceled by $\omega$-factors 
stemming from the numerators in Eq.~(\ref{ctwoex}), as a net 
result, each $1/ \lambda(\nu_n)$ factor reduces 
the order in $\beta E_C$ of the corresponding 
term in the high-temperature expansion of $G_2$ by one order. 
Hence, the quantum correction to $G_2$ of a given order in 
$\beta E_C$ depends  on higher-order terms in the
series expansion of $C_2(\nu_n)$.

An analysis of contributions of all orders \cite{DG} 
shows that a term 
of the series (\ref{exp}) with a product 
\[
S_{2k_1} [\varphi] S_{2k_2} [\varphi] 
\ldots S_{2k_{\ell}} [\varphi]
\]
of quartic or higher-order actions gives quantum corrections 
to $G_2$ of order 
$(\beta E_C)^{k_1+k_2+ \cdots + k_{\ell}- \ell}$ and of higher 
orders. This proves that the terms of the expansion of 
$C_2(\nu_n)$ given explicitly in Eq.~(\ref{ctwoex}) 
suffice to calculate the leading order quantum corrections to $G_2$.

Now, combining Eqs.~(\ref{ctwoex}) and (\ref{gtwo}) one finds
after some algebra the explicit high-temperature result for $G_2$
that may be added to Eq.~(\ref{gone}). This gives
for the conductance
\begin{eqnarray}
G &=&\frac{1}{R_T+R}\bigg\{ 1 \nonumber \\
& & - \frac{R_T}{R_T+R}  
\left[ \frac{\gamma + \psi(1+u)}{u} + \psi^{\prime}(1+u) \right] 
\frac{\beta E_C}{\pi^2} \nonumber\\ 
& &+\ {\cal O} (\beta E_C)^2 \bigg\} \, , \label{result}
\end{eqnarray}
where $\gamma$ is Euler's constant and 
\begin{equation}
u = \frac{g \beta E_C}{2 \pi^2} = 
\frac{R_K(R_T+R)}{R_T R}\, \frac{\beta E_C}{2 \pi^2}\, . \label{udef}
\end{equation}
Further, $\psi(z)$ and $\psi^{\prime}(z)$ denote the 
digamma function and its derivative, respectively. 

\begin{figure}
\epsfysize=9cm
\epsffile{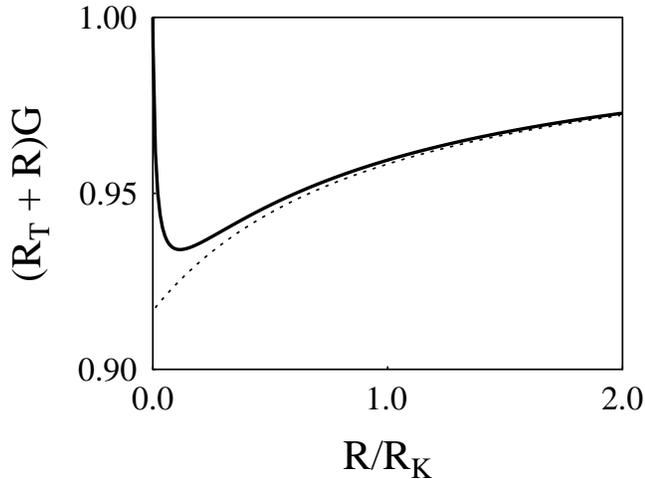}
\caption[]{\label{fig2} The ratio of the conductance $G$ and
the classical conductance $1/(R_T+R)$ is shown vs
the environmental resistance $R$ in units of $R_K$ for
$R_T=R_K$ and $\beta E_C = 1/4$. The full
line shows the result \protect{(\ref{result})}, and the dotted
line depicts the approximation \protect{(\ref{high})} for 
moderate-to-high resistance.}
\end{figure}

The predicted quantum corrections to the classical
conductance $1/(R_T+R)$ reduce the conductance due to the 
Coulomb blockade effect. In Fig.~\ref{fig2} the result (\ref{result})
is depicted for fixed temperature and tunneling resistance
$R_T$ as a function of $R/R_K$.
Typical lead resistances are in the range of 100$\Omega$ so that
$R/R_K$ is usually very small. Then, the parameter $u$
defined in (\ref{udef}) is large even for temperatures $T$ 
well above $E_C/k_B$, and one can use the asymptotic expansion 
of the digamma function to give for $R \ll R_T,R_K$
\begin{equation}
G=\frac{1}{R_T}\left[1 + 2 \frac{R}{R_K} 
\ln \left(\frac{R}{R_K}\right) + 
{\cal O}\left(\frac{ |\ln(\beta E_C)| R}{R_K}\right)\right] \, .
\nonumber
\end{equation}
This shows that for small lead resistance the influence of
the electromagnetic environment cannot be treated as
a weak perturbation since a finite resistance gives
rise to a reduction of $G$ that is non-analytic in $R$. 
This large effect of even a small lead resistance is
clearly seen in Fig.~\ref{fig2}.
On the other hand, for values of $R/R_K$ of order 1 or
larger, $u$ becomes small in the high-temperature limit and
the digamma functions in Eq.~(\ref{result})
may be expanded to yield
\begin{equation}
G = \frac{1}{R_T+R}\left\{ 1 -\frac{R_T}{R_T+R}\,
\frac{\beta E_C}{3} + {\cal O}[(\beta E_C)^2, u\beta E_C]\right\}\, .
\label{high}
\end{equation}
We note that in the limit of weak tunneling, $R_T \gg R, R_K$,
this result  may be approximated further to read
$G = R_T^{-1}[1-\beta E_C/3 + {\cal O}(\beta E_C)^2]$, which is in
accordance with the conventional approach \cite{Devoret,IN}
based on perturbation theory in the tunneling term. Within
this framework higher order terms in $\beta E_C$ can be
evaluated explicitly \cite{EJ}. However, the conventional
approach misses the fact that the quantum corrections to
Ohm's law vanish in the limit $R/R_T \rightarrow \infty$
according to (\ref{high}). For large $R$ many charges
fluctuate across the tunnel junction on the relevant
time scale $RC$ and lead to a suppression of charging effects.

In summary, we have shown that the conductance $G$ of a tunnel junction
is affected by Coulomb charging effects even in the
high-temperature range where $k_BT$ is large compared to the single
electron charging energy $E_C$. In particular, in the experimentally 
relevant range of small environmental impedance $R$, the corrections to
Ohm's law are significant, since $G(R)$ is nonanalytic at $R=0$.
In view of the fact that the conductance can be measured very
accurately, the predicted effects should be readily observable.

The nonclassical behavior of the conductance is also important for
the precise determination of junction parameters from data in the
"classical" regime.  In this context it is important to note that
the result (\ref{result}) remains valid for small tunneling resistance 
$R_T \ll R_K$ as long as $ N R_T \gg R_K$ where 
$N$ is the number of transport channels. Since $N$ is typically 
above 10$^4$ for metallic tunnel junctions, the theory can be
used to determine junction parameters also in the region of 
strong tunneling.

The authors would like to thank Daniel Esteve and Philippe Joyez
for valuable discussions. Financial support was provided by the 
Deutsche Forschungs\-gemein\-schaft (Bonn).

\end{document}